\documentclass[nonacm,sigplan]{acmart}

\usepackage[]{hyperref}

\usepackage[inline]{enumitem}
\setlist{noitemsep,topsep=0pt,parsep=0pt,partopsep=0pt}

\usepackage[subtle]{savetrees}

\usepackage{setspace}
\usepackage[margin=3pt,font={stretch=0.9}]{caption}
\newcommand{\myparagraph}[1]{\smallskip \noindent{\bf {#1}.}}
\newcommand{\out}[1] {}

\newcounter{codeLineCntr}

\reversemarginpar
\newif\ifnotes
\notestrue

\newcommand{\punt}[1]{}

\renewcommand{\eqref}[1]{Equation~(\ref{eq:#1})}

\newcommand{\proc}[1]{\ifmmode\mbox{\textsc{#1}}\else\textsc{#1}\fi}

  \newcommand{\func}[1]{\ifmmode\mathrm{#1}\else\textrm{#1}fi} %
\newcounter{remark}[section]

\newcommand{\projecttitle}{\textsf{qTPU}}

\usepackage{listings}
\usepackage{grumble}
\usepackage{braket}

\begin{document}
% \title{qTPU: A Quantum Tensor Processor Unit}
%\title{qTPU: Hybrid Quantum-Classical Co-Processing using Tensor Networks}
% \title{qTPU: A Quantum Tensor Processing Architecture}
% \title{qTPU: A Hybrid Quantum-Tensor Processing Unit}
\title{Quantum-Classical Computing via Tensor Networks}
% \title{qTPU: Scalable Quantum }
% \title{qTPU: Hybrid Quantum-Classical Tensor Processing}

\author{Nathaniel Tornow}
% \email{nathaniel.tornow@tum.de}
\affiliation{%
  \institution{Technical University of Munich and Leibniz Supercomputing Centre}
  % \city{Munich}
  \country{}
}
\author{Christian B. Mendl}
% \email{christian.mendl@tum.de}
\affiliation{%
  \institution{Technical University of Munich}
  % \city{Munich}
  \country{}
}
% \author{Martin Ruefenacht}
% \email{martin.ruefenacht@lrz.de}
% % \orcid{1234-5678-9012}
% % \author{G.K.M. Tobin}
% % \email{webmaster@marysville-ohio.com}
% \affiliation{%
%   \institution{Leibniz Supercomputing Centre}
%   % \city{Munich}
%   \country{Germany}
% }
\author{Pramod Bhatotia}
% \email{pramod.bhatotia@tum.de}
\affiliation{%
  \institution{Technical University of Munich}
  % \city{Munich}
  \country{}
}

\begin{abstract}
  Circuit knitting offers a promising path to the scalable execution of large quantum circuits by breaking them into smaller sub-circuits whose output is recombined through classical postprocessing. However, current techniques face excessive overhead due to a naive postprocessing method that neglects potential optimizations in the circuit structure. To overcome this, we introduce \projecttitle{}, a framework for scalable hybrid quantum-classical processing using tensor networks. By leveraging our hybrid quantum circuit contraction method, we represent circuit execution as the contraction of a hybrid tensor network (h-TN). The \projecttitle{} compiler automates efficient h-TN generation, optimizing the balance between estimated error and postprocessing overhead, while the \projecttitle{} runtime supports large-scale h-TN contraction using quantum and classical accelerators. Our evaluation shows orders-of-magnitude reductions in postprocessing overhead, a $10^4\times$ speedup in postprocessing, and a 20.7$\times$ reduction in overall runtime compared to the state-of-the-art Qiskit-Addon-Cutting (QAC).

\end{abstract}

\maketitle % should come after the abstract
\pagestyle{plain} % should come right after \maketitle

% \pramod{please populate references.}
\section{Introduction}
Quantum computing could enable exponential speedups in certain computations such as quantum simulation or optimization \cite{kim2023evidence, farhi2014quantum}.
Thereby, an increasing number of data centers are adding quantum processors (QPUs) to their repertoire of computational resources~\cite{IBMQuantum2024, MicrosoftAzureQuantum2024, AmazonBraket2024, GoogleQuantumAI2024}.
However, current QPUs are still subject to technological challenges due to severe noise causing qubit decay and state decoherence, making it currently impossible to outperform alternative classical methods \cite{preskill2018quantum, tindall2024efficient, zhou2020limits}.
% At the same time, the classical simulation of quantum programs is mainly limited by the degree of entanglement, as this is where the computational advantages of QPUs are believed to lie .

To eventually make quantum computing practical, we must employ computations that integrate high-performance classical resources with QPUs, using them as accelerators for specific, suitable tasks \cite{alexeev2024quantum, IBMQuantumRoadmap, elsharkawy2023integration}.
A promising method for scalable hybrid computing is quantum circuit knitting, which facilitates quantum circuit execution by leveraging the strengths of both quantum and classical computing \cite{peng2020simulating, mitarai2021constructing}.
This approach involves decomposing a large quantum circuit into smaller subcircuits, which can be executed with reduced noise on smaller, noisy QPUs. The original circuit's final result is then reconstructed through classical postprocessing (PP).

Theoretically, its divide-and-conquer approach could make circuit knitting a highly suitable mechanism for scaling quantum workloads in data centers with increasingly better QPUs to execute the subcircuits, combined with an abundance of classical resources for the necessary classical PP.
 % suggests that circuit knitting could be a viable strategy for scaling up quantum computations in a hybrid computing environment.

However, the practical applicability of quantum circuit knitting remains greatly hampered by its challenges of prohibitively high exponential sampling and postprocessing (PP) overheads \cite{peng2020simulating, mitarai2021constructing, ufrecht2023optimaljoint, brenner2023optimal, piveteau2022circuit, piveteau2023circuit, tang2021cutqc, Bravyi2016trading, circuit-knitting-toolbox}.
One main limitation of current knitting tools is their reliance on a brute-force PP method, which always assumes worst-case overhead \cite{circuit-knitting-toolbox}. These approaches fail to exploit efficient structures within decomposed circuits that could drastically lower computational costs and hamper the effective use of tensor processing capabilities of GPUs/ TPUs, making it difficult to scale beyond classical techniques.

To overcome these scalability barriers, we pose the following research question:  
\textit{Can we design a circuit knitting-based framework that enables large-scale and GPU-accelerated quantum-classical processing of quantum circuits?}  
Such a framework should be able to outperform classical simulators while producing fewer errors than a single QPU.  

In pursuit of this goal, we identify three key challenges:\newline
\textbf{(1)} We must develop a novel technique that significantly reduces PP overhead by exploiting the structure of decomposed circuits.
\textbf{(2)} We need an automated and scalable procedure to convert large circuits into an optimized format to enable such an efficient technique.
\textbf{(3)} To enable scalable hybrid circuit co-processing, we require a system capable of efficiently executing quantum and classical workloads across a cluster of hybrid resources.

To address these challenges, in this work, we introduce \textit{\projecttitle{}: A framework for large-scale hybrid quantum-classical processing using tensor networks}.
To design such a framework, we make the following core contributions:

\begin{itemize}
    \item We present our core idea of \textbf{hybrid quantum circuit contraction}, which enables us to represent the execution of a quantum circuit as a contraction of a \textit{hybrid tensor network (h-TN)}, essentially accelerating the circuit knitting computation using TNs and accelerating PP by orders of magnitude. (\S~\ref{sec:key_idea})
    \item To enable efficient hybrid circuit contraction and reap the full benefits of our h-TN form, we introduce the \textbf{\projecttitle{} compiler}. The \projecttitle{} compiler converts large quantum circuits into efficient h-TNs, while using hyperparameter optimization to find an optimal, user-defined tradeoff between the expected error of circuits on available noisy QPUs and knitting overheads. To do so, we extend well-studied concepts of optimizing TN contraction paths \cite{gray2021hyper}. (\S~\ref{sec:compiler})
    \item Finally, we present the \textbf{\projecttitle{} runtime}, a system for large-scale contraction of h-TNs using a hybrid cluster of QPUs and GPUs, by building on highly performant tensor processing libraries. The runtime allows for efficient sampling to enable a tradeoff between tolerated error and computational resources. (\S~\ref{sec:runtime})
\end{itemize}

We implement \projecttitle{} in Python, using Qiskit, cotengra, and cuTensorNet \cite{qiskit2024, NVIDIACuTensorNet2024, gray2021hyper}. Our evaluation on an Nvidia A100 GPU shows that, compared to the state-of-the-art framework \textit{Qiskit-Addon-Cutting} \cite{circuit-knitting-toolbox}, we reduce PP overhead by several orders of magnitude. This results in a PP speedup of $10^4$, with an additional GPU speedup of up to 18$\times$. Overall, we achieve an average end-to-end runtime speedup of 20.7$\times$ and reduce PP time to a fraction of the total runtime. 

\section{Background and Motivation}

% We now establish the necessary background for our quantum-classical co-processing approach by using tensor networks (\S~\ref{sec:background:tn}) to efficiently knit large quantum circuits based on quasiprobabilistic decomposition (\S~\ref{sec:background:qpd}).

\begin{figure}[t]
    \centering
    \includegraphics[width=\columnwidth]{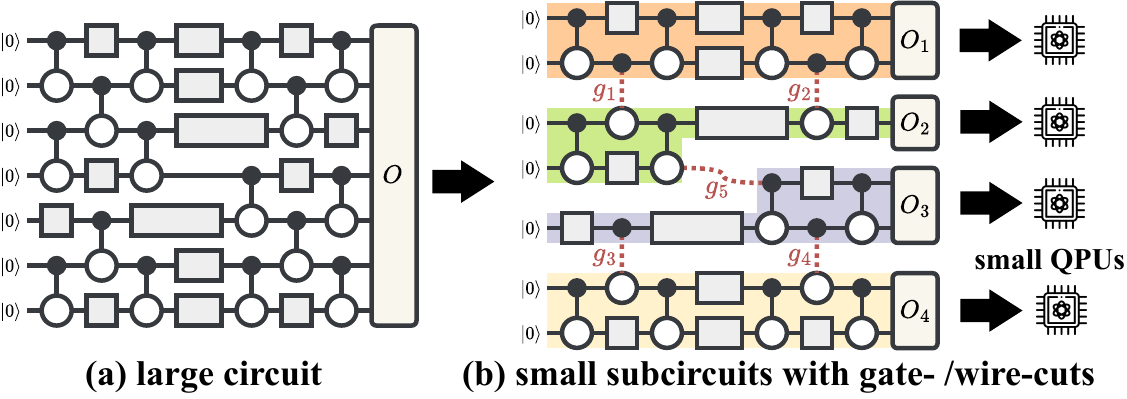}
    \caption{Circuit Knitting. {\em (a) Large quantum circuit. (b) Decomposed quantum circuit with five cuts and four subcircuits that can be run on small QPUs.}}
    \label{fig:background:qpd_cutting}
\end{figure}

\subsection{Challenges of Quantum Circuit Knitting}
\label{sec:back:knitting}
% \pramod{probably shorter it? repeating from the intro.}
Circuit knitting allows us to decompose a large quantum circuit into smaller subcircuits that can be executed independently on smaller QPUs, mitigating noise and scaling circuit execution to more than the size of available QPUs (see Fig.~\ref{sec:back:knitting}) \cite{peng2020simulating, mitarai2021constructing, Bravyi2016trading, tang2021cutqc, circuit-knitting-toolbox, ren2024hardware, yamamoto2022error}. 
Using classical postprocessing (PP), the large circuit's output can be reconstructed with the subcircuit's results, with the final output being an expectation value of an observable $O$.
% or the probability distribution of repeatedly measuring the qubits on a computational basis.
% \pramod{possibly cut and merge references with previous sentence?}
% Previous work has shown that circuit knitting can  \cite{tang2021cutqc, ren2024hardware, yamamoto2022error}.

\begin{figure}[t]
    \centering
    \includegraphics[width=\columnwidth]{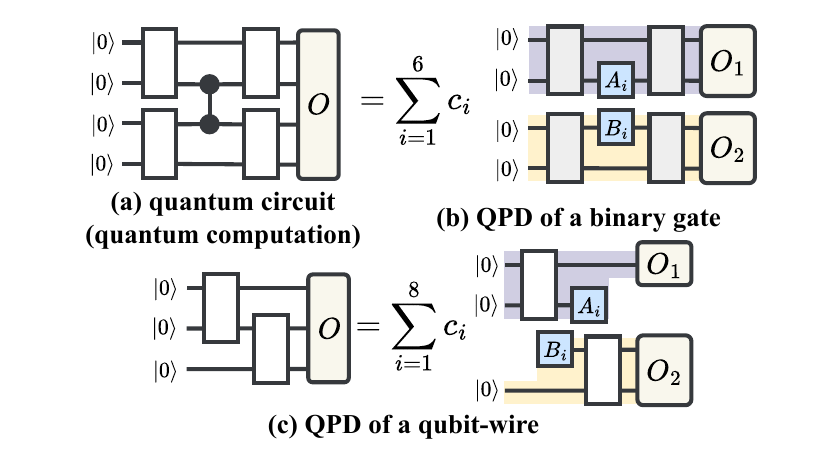}
    \caption{Quasiprobability decomposition (QPD). {\em (a) Quantum computation in the form of a quantum circuit and estimation of the expectation value of the observable $O = O_1 \otimes O_2$. (b) QPD of a two-qubit gate. (c) QPD of a qubit wire.}}
    \label{fig:background:qpd}
\end{figure}

Circuit knitting is primarily built on the \textbf{Quasiprobability Decomposition (QPD)} method, which enables us to decompose qubit wires and multi-qubit gates into a combination of single-qubit unitaries or projective measurements.
The core idea of QPD is depicted in Fig.~\ref{fig:background:qpd}. Here, we cut a two-qubit gate by representing the circuit as a sum of several circuit instances $i$, each weighted by a coefficient $c_i$, where the single-qubit operations $A_i$ and $B_i$ are inserted, replacing the original gate (Fig.~\ref{fig:background:qpd} (b)) \cite{mitarai2021constructing}. Since we only insert single-qubit gates in each instance, we can divide circuits into two smaller subcircuits, which can be executed independently on a smaller QPU.
When assuming that the subcircuits estimate $O_1$ and $O_2$, respectively, with $O = O_1 \otimes O_2$, we can reconstruct the result by calculating $\braket{O} = \sum_{i=0}^{6} c_i \braket{O_1^i} \braket{O_2^i}$ \cite{mitarai2021constructing}.
% While the number of instances $i$ depends on the specific two-qubit gate, the number is $6$ for standard two-qubit gates as $CX$, which we assume in this work.

Similarly, we can cut qubit-wires using QPD (Fig.~\ref{fig:background:qpd} (c)). The only difference to QPD of two-qubit gates is that we need eight instances, and the gates to insert are measurements in different computational bases for $A_i$ and preparation of corresponding states for $B_i$ \cite{peng2020simulating}.

To generalize the circuit knitting to multiple cuts in a circuit as in Fig.~\ref{fig:background:qpd_cutting} (b), we define a coefficient vector $\mathbf{c}_{g_k} = (c_1, c_2, \dots)$ for each decomposed gate $g_k$, and a \textit{global} coefficient vector $\mathbf{C} = \bigotimes_{g_k} \mathbf{c}_{g_k}$.
To then reconstruct the result $\braket{O}$ with circuit knitting, we need to calculate
\begin{equation}
    \braket{O} = \sum_{c_i \in \mathbf{C}} c_i\ \prod_{j=1}^s \braket{O_j^i} ,
\label{eq:knitting}
\end{equation}
where $s$ is the number of subcircuits and $\braket{O_j^i}$ the result of the $j$th subcircuit in the $i$th \textit{global} instance.

% \subsubsection*{Circuit Knitting Overheads}
Severe overheads greatly limit the scalability of QPD-based circuit knitting \cite{mitarai2021constructing, peng2020simulating}. These overheads can be categorized into (1) the sampling and (2) the PP overhead:

The \textbf{sampling overhead} is the number of samples/shots we must draw from the circuit's instances to estimate the result to a given precision.
The sampling overhead is given as $\mathcal{O}(9^k / \varepsilon^2)$ and $\mathcal{O}(16^k / \varepsilon^2)$ total samples for reconstructing the output with a precision $\varepsilon$ for $k$ gate/wire cuts \cite{mitarai2021constructing, peng2020simulating}. 
% \pramod{the following can be cut or merged with the related work.}

% Therefore, the sampling overhead is independent and not the focus of this work.

The \textbf{postprocessing (PP) overhead} is the overhead that stems from the classical PP of the subcircuit's results. We define the PP overhead as the number of float-multiplications (FLOPs) required to perform the PP following Eq.~\ref{eq:knitting}.
If we use a \textit{naive} implementation of Eq.~\ref{eq:knitting}, as currently implemented \cite{circuit-knitting-toolbox}, we require $|\mathbf{C}| (s + n_g - 1)$ FLOPs to postprocess the results, where $n_g$ is the number of QPD-gates/-wires \cite{circuit-knitting-toolbox}, as we need $|\mathbf{C}| (n_g - 1)$ FLOPs to compute $\mathbf{C}$ and $|\mathbf{C}| \cdot s$ FLOPs to compute Eq.~\ref{eq:knitting}.
We can assume that the number of subcircuits $s$ increases sub-linearly and $|C|$ increases exponentially with the number of cuts.
The PP overhead is therefore bounded by $\mathcal{O}(|\mathbf{C}|) \subseteq \mathcal{O}(8^k)$ when cutting $k$ gates or wires.

% \begin{table}[h]
% \centering
% % \renewcommand{\arraystretch}{1.5} % Adjust row height
% \begin{tabular}{|c|c|c|}
% \hline
% \textbf{Type}    & \textbf{Sampling Overhead}       & \textbf{Postproc. Cost}   \\ \hline
% CX-Gate & $\mathcal{O}\left(9^k / \epsilon^2\right)$   & $\mathcal{O}\left(6^k\right)$   \\ \hline
% $R_{ZZ}(\pi)$  & $\mathcal{O}\left(9^k / \epsilon^2\right)$   & $\mathcal{O}\left(6^k\right)$   \\ \hline
% $U$  & $\mathcal{O}\left(9^k / \epsilon^2\right)$   & $\mathcal{O}\left(6^k\right)$   \\ \hline
% Wire      & $\mathcal{O}\left(16^k / \epsilon^2 \right)$  & $\mathcal{O}\left(4^k\right)$ \\ \hline
% \end{tabular}
% % \caption{Asymptotic sampling overheads and postprocessing costs for cutting $k$ gates or wires and reconstructing the result up to an error of $\epsilon$.}
% \end{table}

\subsubsection*{Problem: Brute-force postprocessing} 
State-of-the-art circuit knitting methods face a significant challenge due to an unnecessary exponential overhead arising from a naive implementation, failing to account for efficient structures that could reduce this overhead. To highlight the limitations of the current naive method, we use the decomposed circuit in Fig.~\ref{fig:background:qpd_cutting} (b) as an illustrative example.
Here, we use four gate-cuts ($g_{1\text{-}4}$) and one wire-cut ($g_5$) to decompose the circuit into $s=4$ subcircuits.
Here, the total PP overhead is 
$|\mathbf{C}| (s + n_g - 1) = 10368 \cdot (4 + 5 - 1) = 82944$ FLOPs, with an asymptotic PP cost of 
$\mathcal{O}(|\mathbf{C}|)$ where $ |\mathbf{C}| = 10368$ subcircuit instances must be generated.

However, the structure of the decomposed circuit suggests that a significantly more efficient approach to knitting the circuit may be possible. In particular, the circuit structure is relatively sparse; for instance, subcircuits 1 and 2 can be knitted completely independently from subcircuit 4. Additionally, QPD gates $g_1$ and $g_2$ only affect subcircuits 1 and 2, leaving subcircuits 3 and 4 unaffected. Thus, creating instances for each global $c_i \in \mathbf{C}$ is unnecessary.

Despite this potential for efficiency, the current naive PP approach still requires us to compute a sum over the entire tensor product of the coefficients $\mathbf{C}$. As a result, we are bound to the worst-case PP overhead, regardless of how sparse the circuit structure may be. Furthermore, the naive implementation of a sum over every global coefficient complicates efforts to speed up PP using parallel processing on GPUs \cite{circuit-knitting-toolbox}.

\begin{figure}
    \centering
    \includegraphics[width=.85\columnwidth]{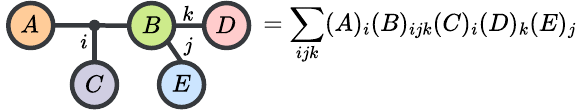}
    \caption{Tensor network (TN) (\S~\ref{sec:background:tn}). 
    {\em A tensor network of five tensors, connected by two indices and one hyper-index, depicted by a hyperedge connecting the three tensors $A$, $B$, and $C$. The sum on the right represents the contraction of the tensor network, resulting in a single scalar.}}
    \label{fig:tn}
\end{figure}

\subsection{Opportunities of Tensor Networks}
\label{sec:background:tn}

Tensor networks (TNs) enable the efficient storage and manipulation of high-dimensional data, making them a powerful tool for quantum circuit simulation \cite{orus2019tensor, bridgeman2017hand, orus2014practical, cichocki2014tensor}. A TN provides a graphical representation of the contraction of multiple tensors. In this representation, vertices represent tensors, and edges represent the indices of those tensors. 
A contraction occurs across an index when two or more tensors are connected via an edge or hyper-edge. Fig.~\ref{fig:tn} illustrates a TN with five tensors labeled $A$ to $E$ and three indices. When indices connect multiple tensors, we contract by "summing over" the corresponding indices, as indicated by the expression on the right.

The primary reason for the effectiveness of TNs is their ability to efficiently represent quantum states of large systems that exhibit low entanglement \cite{fannes1992finitely, hastings2007}. A quantum circuit execution can also be represented as an equivalent TN, which can be contracted to obtain the result \cite{zhou2020limits, gray2018quimb, seitz2023simulatingquantum, tindall2024efficient, NVIDIACuTensorNet2024}. 
However, the main limitation of this approach is the degree of intra-connectivity between the gates or qubits within the circuit. Higher connectivity tends to increase the complexity of the tensor contractions, impacting performance.

\begin{figure}
    \centering
    \includegraphics[width=.8\columnwidth]{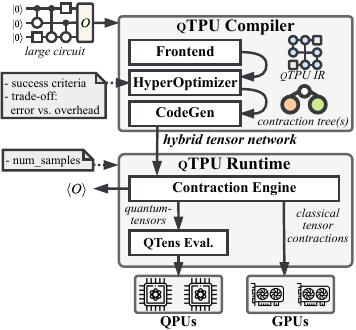}
    \caption{\projecttitle{} System Overview (\S~\ref{sec:overview}).}
    \label{fig:overview}
\end{figure}

% \pramod{i think the next two paragraphs can be merged to save space.}

% \pramod{repeating the point in the first para of this section.} 
% TNs have been proven as a prominent mechanism for efficiently simulating quantum circuits \cite{gray2018quimb, gray2021hyper, NVIDIACuTensorNet2024}. 
% This means we can efficiently simulate wide or deep quantum circuits that exhibit low entanglement \cite{}.

% \pramod{i believe you are using compression and contraction interchangeably!}
\subsubsection*{Motivation: Accelerating circuit knitting}
In this work, we contend that TNs could also effectively reduce PP overhead by accelerating computation during this phase. 
% Notably, we observe that naive knitting (as expressed in Eq.~\ref{eq:knitting}) is itself a form of tensor contraction. 
Thus, we aim to express circuit knitting as an efficient TN, which can significantly decrease PP costs for suitable circuit cuts. By leveraging TN processing for PP, we can take advantage of existing, highly optimized tensor-processing libraries that allow us to offload computations to GPUs, further enhancing performance.

\section{Overview}
\label{sec:overview}

\begin{figure*}[t]
    \centering
    \includegraphics[width=\textwidth]{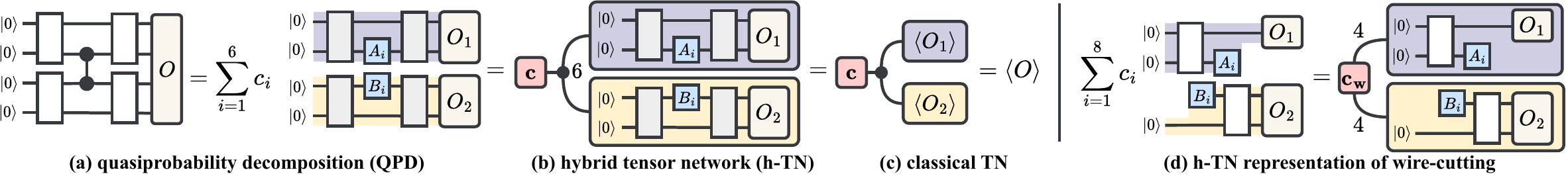}
    \vspace{-3mm}
    \caption{Hybrid Quantum Circuit Contraction (\S~\ref{sec:key_idea}). {\em Through Quasiprobability-Decomposition (QPD), we can represent the execution of a quantum circuit as a hybrid tensor network (h-TN) of both quantum and classical tensors.}}
    \vspace{-3mm}
    \label{fig:key_idea}
\end{figure*}

% \subsection{Limitations of Circuit Knitting Approaches}

% CutQC and CKT

% s\pramod{data center scale, instead of HPC scale?}
We aim to develop a data center-scale co-processing framework that enables efficient, hybrid execution of large quantum circuits using QPUs and classical tensor accelerators.
To do so, we identify three core design challenges.

\subsection{Design Challenges}

\subsubsection*{Challenge \#1: Enabling quantum-classical processing}
As pointed out in \S~\ref{sec:back:knitting}, PP of QPD-based circuit knitting suffers from severe exponential overheads, which in many cases, depending on the specific structure of the cuts, can be unnecessarily high due to the currently used naive approach.
To enable circuit knitting as a viable hybrid quantum-classical processing technique that could go beyond a handful of circuit cuts, we need a new approach similar to the approach of TNs in quantum simulation that reduces the PP overhead to an acceptable level.

\subsubsection*{Challenge \#2: Efficiency and adaptability}
To fully utilize a scalable hybrid processing technique, we need automatic and efficient procedures to transform a quantum circuit into an optimized quantum-classical program. This optimized program should offer an optimal trade-off between (1) PP overhead and (2) other user-specified overheads, such as the error incurred when executing subcircuits on noisy QPUs.
In the context of circuit knitting-based processing, this requires identifying efficient circuit partitions that balance the dual objectives of minimizing both quantum and classical overhead while accounting for the expected noise from running subcircuits. Ideally, these circuit-cutting techniques should be near-optimal and capable of scaling to handle large circuits with thousands of qubits and significant depths.

\subsubsection*{Challenge \#3: Large-scale hybrid processing}
Finally, to execute the hybrid quantum-classical workflow at scale, we need a system that supports operation on a hybrid cluster of multiple QPUs and GPUs. The challenges in this context include adapting the computational cost through sampling and efficiently offloading quantum and classical tasks to their respective devices using parallel processing. Additionally, we should be able to leverage existing, highly optimized tensor-computing libraries to enhance performance.

\subsection{The \projecttitle{} Framework}

To tackle the challenges outlined above, we present \projecttitle{}, a framework for large-scale quantum-classical co-processing (see Fig.~\ref{fig:overview}). \projecttitle{} is built upon our core concept of \textit{hybrid quantum circuit contraction} (see Fig.~\ref{fig:key_idea}). To execute a large circuit across a set of QPUs and GPUs, we divide the framework into two components: the \textit{\projecttitle{} compiler} and the \textit{runtime}.
% We show an overview of the \projecttitle{} framework in Fig.~\ref{fig:overview}.

\myparagraph{Hybrid quantum circuit contraction}
Hybrid circuit contraction is our approach to minimizing the PP overhead associated with knitting a given circuit. The core idea is to represent the execution of a quantum circuit as a \textbf{hybrid tensor network (h-TN)} that consists of quantum and classical tensors. 
By encoding the execution as an h-TN, we effectively represent the knitting procedure in a more efficient format based on a TN. This technique enables a best-case PP overhead that can outperform current PP-methods by several orders of magnitude, depending on how the original circuit is compiled.
% We describe the core idea in detail in \S~\ref{sec:key_idea}.

\begin{figure*}
    \centering
    \includegraphics[width=\textwidth]{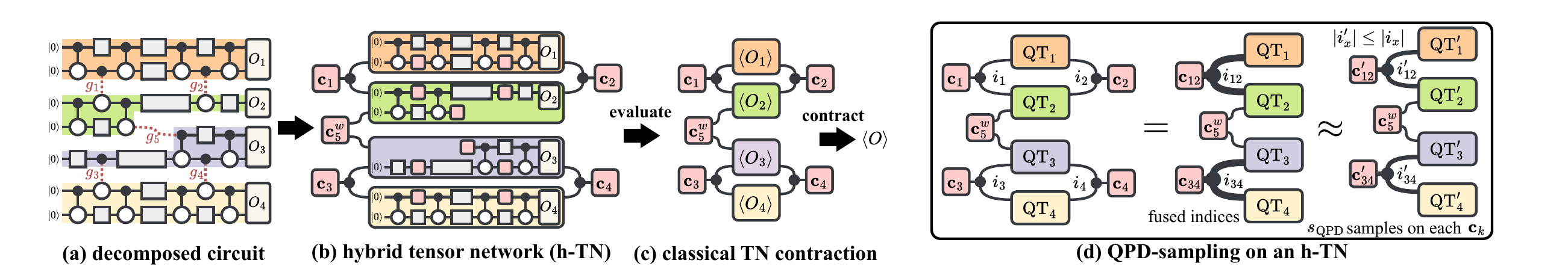}
    \vspace{-3mm}
    \caption{Contraction of a hybrid tensor network (h-TN) (\S~\ref{sec:key_idea}). {\em (a) The decomposed circuit of Fig.~\ref{fig:background:qpd_cutting}. (b) The corresponding h-TN. (c) the evaluated (classical) TN that contracts to the result $\braket{O}$. (d) QPD-sampling as a simplification of the h-TN.}}
    \vspace{-3mm}
    \label{fig:tn_knit}
\end{figure*}

\myparagraph{\projecttitle{} compiler}
As the performance of our hybrid circuit contraction technique depends on the specific structure of a circuit's decomposition, we present the \projecttitle{} compiler, which compiles a large quantum circuit into an efficient h-TN while finding a balance between classical overhead and errors caused by QPUs.
To do so, the compiler consists of three main components:
First, the \textit{frontend} converts the circuit into the \projecttitle{} intermediate representation (IR), a TN-like representation of the circuit.
Then, the \textit{optimizer} finds an optimal decomposition of the IR using contraction trees. The optimization is based on user-given success criteria and a trades off between expected error and processing overheads.
Finally, the \textit{code generator} transforms the decomposed IR into an h-TN.
% We describe the compiler in detail in \S~\ref{sec:compiler}.

\myparagraph{\projecttitle{} runtime}
For the large-scale contraction of an h-TN, we present the \projecttitle{} runtime.
To contract an h-TN, the runtime first applies sampling to reduce the requirements of classical memory and time resources at the cost of approximations.
Then, the \textit{contraction engine} facilitates the hybrid contraction by dividing the contraction into quantum and classical tasks.
For the quantum-part of the contraction, quantum-tensors are evaluated by running quantum circuits on a set of QPUs using the \textit{quantum-tensor evaluator}.

\section{Hybrid Quantum Circuit Contraction}
\label{sec:key_idea}

% To enable HPC-scale and GPU-based hybrid co-processing through circuit knitting, we must significantly reduce the overheads compared to current brute-force knitting methods. 
% \pramod{maybe the abstraction/design should be independent of GPUs -- we should maybe frame it for acceleration. Also we could pitch it as a new abstraction here. Repeat the overarching story.}
Our goal is to enable large-scale and accelerated hybrid processing, which is currently unfeasible in circuit knitting due to its naive approach. To address this, we introduce our key concept of \textit{hybrid circuit contraction}, which represents circuit knitting in the form of a \textit{hybrid tensor network (h-TN)}.

\subsection{Key Idea: Hybrid Quantum-Tensor Contraction}
\label{sec:key_idea:hybrid_contraction}

To grasp the concept behind hybrid circuit contraction, we first need to understand how to represent a QPD of a single gate or wire as an h-TN.
We illustrate this relationship in Fig.~\ref{fig:key_idea}. Here, Fig.~\ref{fig:key_idea} (a) shows the standard QPD formula as a sum over 6 circuit instances $i$, each decomposed into two subcircuits and weighted by coefficients $c_i$.
We can then express this formula as an h-TN, as shown in Fig.~\ref{fig:key_idea} (b). To achieve this, we group the coefficients into a classical tensor (CT) $\mathbf{c} = (c_1, \dots, c_6)$ and the instances of the two subcircuits into two quantum tensors (QTs). A QT is essentially a tensor (in this case, a vector), with its elements representing the instances of the respective subcircuit. To match the QPD formula, the tensors are connected via a hyper-index.

To contract an h-TN, we first \textit{evaluate} each QT by running the instances that yield the outputs $\braket{O_j^i}$ for the $j$th subcircuit in the $i$th instance. For each QT, the results are written into a corresponding classical tensor (CT) $\braket{O_j}$, where the elements represent the results of the respective instances (as shown in Fig.~\ref{fig:key_idea}(c)).  
Finally, we \textit{contract} this fully classical TN to obtain the final result, $\braket{O}$.

% We present our key idea of hybrid quantum-classical circuit contraction in Fig.~\ref{fig:key_idea}. In (a), we show the QPD method as described in \S~\ref{sec:background:qpd} expressed as the weighted sum over the circuit instances $i$. In each instance $i$ we multiply the results of the subcircuits $\braket{O_1}_i$ and $\braket{O_2}_i$.

% We now propose to represent the QPD equation as a contraction of a \textit{hybrid TNk} (h-TN), as shown in Fig.~\ref{fig:key_idea} (b). This h-TN consists of two \textit{quantum tensors} and one \textit{classical tensor}.
% In particular, a quantum tensor can be considered the tensor of multiple quantum circuits, where the two quantum tensors are the subcircuits in their respective instances.
% Both quantum tensors are connected to the classical tensor $\mathbf{c} = (c_1, ..., c_n)$ via a hyperindex.

% To contract this h-TN, we first need to \textit{evaluate} the quantum tensors. To do so, we run each circuit of the quantum tensor and write the respective results $\braket{O_j}_i$ into a classical tensor, as depicted in Fig.~\ref{fig:key_idea} (c). This now entirely classical TNk's contraction gives the result $\braket{O}$.

Similarly, to represent a wire cut as an h-TN, we can apply the same method using the identity shown in Fig.~\ref{fig:background:qpd}(c). However, to be more efficient, we can adopt the approach from CutQC \cite{tang2021cutqc}, which performs a wire cut using only four basis measurements $A_i$ and four state preparations $B_i$, as depicted in Fig.~\ref{fig:key_idea}(d).
Unlike the h-TN for a gate, this approach requires a two-dimensional $4 \times 4$ CT and four instances $A$ and $B$ for the two QTs, following the CutQC method closely:
\begin{equation}
    c_w = \frac{1}{2} \begin{pmatrix}
        1  & 1  & 0 & 0 \\
        1  & -1 & 0 & 0 \\
        -1 & -1 & 2 & 0 \\
        -1 & -1 & 0 & 2
    \end{pmatrix}\ \
    \begin{array}{c}
        A = (I,\ Z,\ X,\ Y), \\
        B = (\ket{0},\ \ket{1},\ \ket{+},\ \ket{i})
    \end{array}
    \label{eq:wire-tensor}
\end{equation}
This approach drastically reduces the memory complexity and contraction cost with multiple cuts.

\subsection{Postprocessing Overhead Analysis}
Through hybrid circuit contraction, we can now represent individual circuit cuts as an h-TN. Next, we analyze the impact of this approach on PP overhead when cutting multiple gates or wires, as illustrated by the example in Fig.~\ref{fig:tn_knit}.

Here, we consider the same decomposed circuit with 5 cuts that we discussed for the current brute-force knitting in \S~\ref{sec:back:knitting}. Using our method, this circuit can now be represented as an h-TN, consisting of five classical tensors (CTs), one for each QPD, four quantum tensors (QTs), and one for each subcircuit.
% For instance, the top and bottom QTs have dimensions of $6 \times 6$, while the other QTs have dimensions of $4 \times 6 \times 6$.

To contract the h-TN, we first evaluate each quantum tensor, yielding the corresponding classical tensors building a classical TN in Fig.~\ref{fig:tn_knit} (c).
The classical PP then only consists of contracting the classical TN. The (optimal) cost of contracting the TN of Fig.~\ref{fig:tn_knit} (c) is $884$ FLOPs, which is \textit{orders of magnitude} smaller than the cost of the brute-force knitting approach of \S~\ref{sec:back:knitting}.
We obtain this cost by first contracting the tensors of $\braket{O_1}$ and $\braket{O_2}$, as well as $\braket{O_3}$ and $\braket{O_4}$ with their respective connecting classical tensors $\mathbf{c}_{1-4}$, each with a cost of 432 FLOPs. Finally we contract the two resulting tensors and the $\mathbf{c}_5^w$ tensor with 20 FLOPs, resulting in a total of 884 FLOPs.

The PP overhead of hybrid circuit contraction relies entirely on the contraction cost of the final classical TN. This cost heavily depends on the order of tensor contractions performed, making an efficient contraction sequence essential \cite{gray2021hyper, NVIDIACuTensorNet2024}. 

In the worst case, contracting a TN is asymptotically equivalent to the current knitting method described in Eq.~\ref{eq:knitting} when we fail to identify a more efficient contraction sequence or when the h-TN structure does not allow for it \cite{bridgeman2017hand}. Conversely, in the best case, if the h-TN includes sufficient QTs connected by CTs and we find an optimal contraction sequence, we achieve minimal PP overhead for a given decomposed circuit.

The effectiveness of our approach in reducing PP costs depends on our ability to accelerate the knitting computation by leveraging the sparse structure of the decomposed circuit, akin to existing TN methods for efficient representations of large quantum states \cite{fannes1992finitely, vidal_efficient_2003}. This approach occurs by treating the knitting process as a series of pairwise independent knitting operations between subcircuits rather than knitting the entire circuit in one step. For example, subcircuits 1 and 2, as well as subcircuits 3 and 4, are knitted independently through efficient partial contractions.

\begin{figure*}[t]
    \centering
    \includegraphics[width=\textwidth]{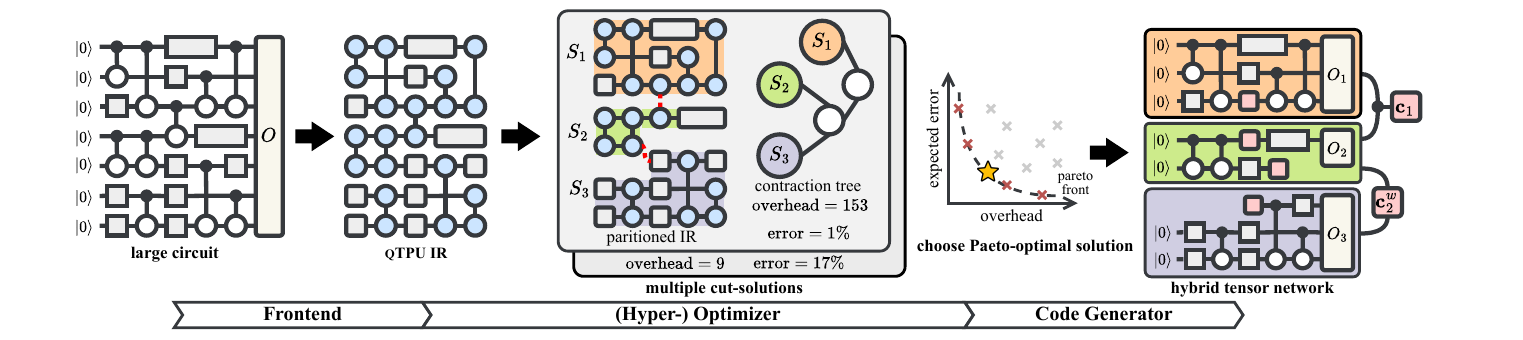}
    \vspace{-5mm}
    \caption{Workflow of the \projecttitle{} Compiler. {\em First, the frontend generates the \projecttitle{} IR. Then, the hyperoptimizer optimizes the IR using multiobjective optimization between contraction cost and success probability, producing multiple Pareto-optimal solutions, each with a contraction tree that describes a partition of the IR. The code generator generates the corresponding h-TN of user-selected solution.}}
    \vspace{-3mm}
    \label{fig:compiler:workflow}
\end{figure*}

\subsection{QPD-Sampling and h-TN Simplification}
\label{sec:idea:approx}

In addition to its postprocessing (PP) overhead, circuit knitting incurs a significant sampling overhead that determines the total number of samples required to estimate the expectation value $\braket{O}$ to an accuracy $\varepsilon$ \cite{mitarai2021constructing, peng2020simulating} (\S~\ref{sec:back:knitting}). We can reduce the number of samples $s_{\text{QPD}}$ to lessen the overhead, but this comes at the cost of lower-quality approximations.

In the current brute-force knitting approach, we sample $s_{\text{QPD}}$ instances and assign them to global circuit instances $i$, each weighted by its respective coefficient $|c_i|$, where all $c_i$ belong to $\mathbf{C} = \bigotimes \mathbf{c}_k$, the tensor product of all coefficients of QPD gates (see \S~\ref{sec:back:knitting}) \cite{mitarai2021constructing}. Instances with not-sampled coefficients can be ignored, enabling faster knitting. This process reduces the upper bound of PP to $\mathcal{O}(s_{\text{QPD}})$ \cite{circuit-knitting-toolbox}.

In contrast, using our hybrid circuit contraction, we can perform equivalent sampling directly in the form of the h-TN, which still gives an exponential reduction in PP cost, as shown in Fig.~\ref{fig:tn_knit} (d).
To do so, we first simplify the h-TN by fusing the indices between every two QTs by computing the tensor product of the CTs between the QTs, giving an equivalent but more compact form of the h-TN.
Then, we perform sampling on the individual CTs $\mathbf{c}_x$, assigning samples to $(\mathbf{c}_x)_i$ and weighting by $|(\mathbf{c}_x)_i|$.
The samples we record along an index determine the number of shots each circuit instance of an adjacent QT must be performed.
For all $(\mathbf{c}_x)_i$ that are not sampled at all, we can truncate the CT and its respective index $i_x$, as we sketch on the right in Fig.~\ref{fig:tn_knit} (d).
This truncation also allows us to reduce the size of the respective QTs along the given index, further reducing PP overhead.

Note that the sampling does not apply to the unique wire-cut tensor (Eq.~\ref{eq:wire-tensor}), as all instances are equally weighted \cite{circuit-knitting-toolbox}.

\section{\projecttitle{} Compiler}
\label{sec:compiler}

% In \S~\ref{sec:key_idea} we presented a technique that enables scalable hybrid quantum-classical co-processing using our hybrid circuit contraction. To take full advantage of our approach, we need a method that automatically finds gates or wires that need to be cut. This method minimizes the cost of post-processing, considering that post-processing is now an efficient tensor network contraction, and minimizes the possible noise in the execution of the respective QTs on QPUs. We tackle this challenge by introducing the \projecttitle{} compiler, a pipeline to transform a quantum circuit into an efficient h-TN, minimizing errors of running subcircuits on noisy QPUs as well as post-processing cost.

% To find an efficient hybrid tensor network that allows for a tradeoff between processing cost and high-fidelity computation for an arbitrary quantum circuit, we present the \projecttitle{} compiler.

% To enable the compilation of quantum circuits into efficient hybrid tensor networks (h-TN), we present the \projecttitle{} compiler. 

% \subsection{Workflow}
% \label{sec:compiler:workflow}

To fully leverage our hybrid circuit contraction method (\S~\ref{sec:key_idea}), we need an efficient process to transform large circuits into optimal h-TNs, balancing contraction overhead and estimated error. We address this with the \projecttitle{} compiler, as shown in Fig.~\ref{fig:compiler:workflow}.

The process starts with the frontend converting the circuit into the \projecttitle{} IR, a TN-like graph of the circuit’s decomposition. The optimizer then applies hyperparameter optimization, generating Pareto-optimal solutions that trade off estimated error and overhead. The error metric can be user-defined, and could range from QPU noise error to error correction overhead.
Each solution comprises a partitioned IR and a contraction tree, which dictates the hybrid TN's overhead. We select one of the optimal solutions based on the user's preference and finally generate h-TN with its quantum and classical tensors.

% The \projecttitle{} compiler converts a large quantum circuit into an efficient hybrid tensor network (h-TN) and optimized contraction sequence. To do so, the compiler follows the three main steps of (1) the frontend, (2) the optimizer, and (3) the code-generator, as we illustrate in Fig.~\ref{fig:compiler-workflow}.

% First, the compiler \textit{frontend} converts the circuit into the representation of a circuit graph IR (CG-IR).
% Based on the CG-IR, an \textit{optimizer} computes an efficient graph-partitioning and a corresponding optimized knit-tree of the CG.
% Finally, the \textit{code-generator} generates the h-TN by generating the QTs with their respective instantiable quantum circuits and the classical tensors for the respective QPD gates. For the contraction of the h-TN, we also compute an optimized contraction order based on the classical device that will perform the final contraction of the h-TN.

\subsection{Frontend and \projecttitle{} IR}
\label{sec:compiler:frontend}

The compiler frontend converts a quantum circuit into the \projecttitle{} IR, which captures the connectivity of circuit operations, allowing for efficient optimization and h-TN generation.

In particular, the IR is a graph consisting of pairs $(\text{op}_i, q_x)$ as vertices, where $\text{op}_i$ is the operation-index, and $q_x$ a qubit in the circuit. 
Gate-edges (vertical) exist for two-qubit gates between edges having the same operation $\text{op}_i$, and wire-edges (horizontal) exist between subsequent operations for vertices having the same qubit $q_x$.
% There exists a (gate-) edge between $(\text{op}_i, q_x)$ and $(\text{op}_i, q_y)$ if the two-qubit $\text{op}_i$ is acting on qubits $q_x$ and $q_y$ (horizontal edges). A (wire-) edge exists between $(\text{op}_i, q_x)$ and $(\text{op}_j, q_x)$ if $\text{op}_j$
% follows $\text{op}_i$ on qubit $q_x$ (vertical edges).
Each edge is weighted by the sampling overhead of the respective gate/wire to guide the optimization.
% For unparameterized gates and wires, this weight is fixed (e.g., $w_{CX} = 9$ and $w_{wire} = 16$), whereas, for parameterized gates such as gates, the overhead depends on the parameter (e.g., $w_{RZZ(0)} = 1$ and $w_{RZZ(\pi/2)} = 9$).
Thus, we aim to cut the least amount of gates/wires that have the least amount of sampling overhead, such that we (1) minimize the PP overhead and (2) enable the highest amount of overhead reduction during sampling (\S~\ref{sec:idea:approx}).

\begin{figure}[t]
    \centering
    \includegraphics[width=\columnwidth]{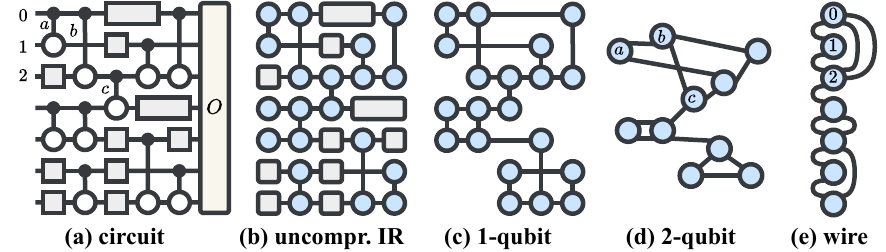}
    \caption{Compression of the \projecttitle{} IR. {\em }}
    \label{fig:compiler:compression}
\end{figure}

% Note that we choose the representation of the IR as it closely resembles the TN representation of a quantum circuit \cite{}.
% Therefore, this representation enables efficiently utilization of existing, well-studied tools for optimizing TN contraction sequences.

% Therefore, during optimization, we can reliably determine a possible solution's overall (post-) processing cost.
% \nate{explain why we don't use the circuit overhead}

\subsubsection*{IR compression}
To enable more efficient optimization, we compress the IR, as illustrated in Fig.~\ref{fig:compiler:compression}. Each node in a compressed graph represents a subgraph of nodes from the uncompressed graph. We implement three main compression techniques: One-qubit compression merges single-qubit vertices into neighboring two-qubit operation vertices (Fig.~\ref{fig:compiler:compression} (c)). Two-qubit compression merges each vertex of the same operation (Fig.~\ref{fig:compiler:compression} (d)). Wire compression merges all vertices corresponding to the same qubit-wire (Fig.~\ref{fig:compiler:compression} (e)).

Since single-qubit gates contribute minimally to errors compared to two-qubit operations \cite{IBMQuantum2024}, one-qubit compression reduces the search space without affecting our ability to find an optimal solution. In contrast, two-qubit and wire compression restrict the search space more aggressively but in a controlled manner: two-qubit compression focuses the search on wire cuts, while wire compression restricts it to gate cuts. This allows users to set compression as a parameter to prioritize wire or gate cuts or to use it as a hyperparameter to accelerate the search in the desired subspace.

\begin{figure*}[t]
    \centering
    \includegraphics[width=\textwidth]{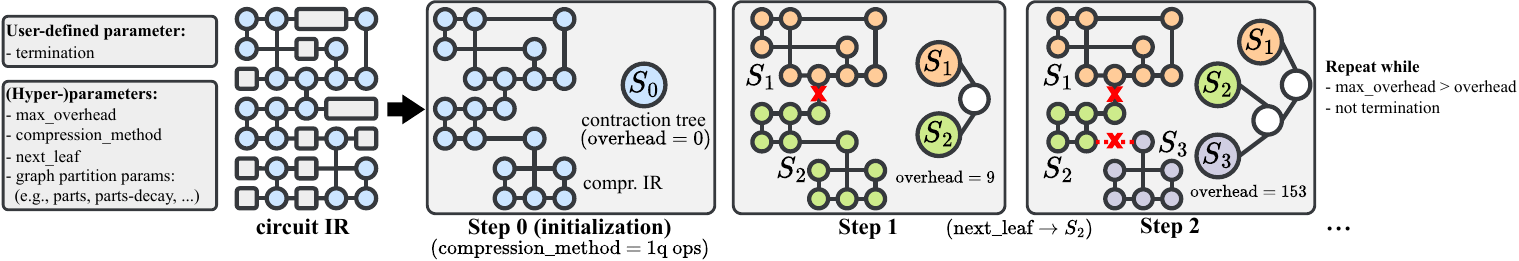}
    \vspace{-5mm}
    \caption{Optimizer. {\em Based on a set of user-defined hyperparameters, we optimize the circuit IR by first applying compression and then iteratively building a contraction tree by (greedily) choosing a tree-leaf to partition until we reach a maximum cost or we reach a terminating condition.}}
    \vspace{-3mm}
    \label{fig:optimizer}
\end{figure*}

% The front end converts a quantum circuit into the circuit graph representation.
% In the circuit graph $G_C = (V, E)$ the vertices are tuples $(\text{op}_i,\ q_j) \in V$ of a operation-id $\text{op}_i$ and a qubit $q_j$.

% So, an edge represents an operation when its two vertices share the same $\text{op}_i$, and a qubit-wire when they share the same qubit $q_j$.

% To guide the following optimization of the circuit graph through partitioning, each edge is assigned a weight $w$, corresponding to the number of instances $n$ needed if we apply QPD to the wire or operation of the respective edge.

% \subsubsection{Knit Tree}
% An optimizer uses the knit-tree data structure to optimize a circuit graph.
% A knit tree is a binary tree where each node recursively represents a subset of nodes in the circuit graph, meaning that a node represents the union of the subsets of its two children.

\subsection{Optimizer}
\label{sec:compiler:optimizer}

In the second stage of the compiler, the optimizer aims to generate an optimal partition of the IR using a contraction tree.
In doing so, it aims to minimize the expected errors of running the resulting subcircuits on (noisy) QPUs and overhead while reaching a user-defined success criterion.

% We use \textit{estimated probability of success (EPS)} (or estimated fidelity) as a measure of how likely it is that an h-TN can be successfully executed, especially depending on the hardware limitations of the QPUs.
% While we allow users to specify their own measure of EPS, by default the compiler uses a simple uniform error model, e.g., with an error of $e = 10^{-4}$ for each one-qubit and an error of $e = 10^{-3}$ for each two-qubit gate.
% For each subcircuit $s$ in the partitioned IR, we assign $\text{EPS}_s = \prod_i (1-e_k)$, where $e_k$ is the error of the $k$th gate, and choose the minimum $\text{EPS}_s$ overall.

\subsubsection*{Contraction tree}
A contraction tree (Fig.~\ref{fig:compiler:workflow}, center) is a binary tree for recursively determining the asymptotic overhead of the \projecttitle{} IR \cite{gray2021hyper}. In particular, a leaf in the contraction tree describes a subset of vertices in the IR, and each non-leaf vertex describes the contraction between two subsets.
In the example of Fig.~\ref{fig:compiler:workflow}, the contraction tree determines that we must contract $S_2$ and $S_3$ and then contract the result with $S_1$.

\subsubsection*{Hyperparameters}
The optimization workflow depends on the following hyper-parameters:
% , which can be hyperparameters for the hyperparameter optimization or chosen by the user.

\begin{itemize}
    \item \textit{Termination}: A function to terminate the optimization. E.g., if the qubit-number is below a given threshold.
    \item \textit{Max Overhead}: The maximum (asymptotical) PP overhead.
    \item \textit{Compression}: A \projecttitle{} IR compression method to speed up optimization or guide the optimization into the search space of either only wire- or gate-cuts (\S~\ref{sec:compiler:frontend}).
    \item \textit{Next-Leaf}: A function determining the leaf of the contraction tree to partition next. 
    % We implement greedy functions that choose the leaf with the maximum qubits or vertices respectively.
    \item \textit{Graph Partition Params}: Parameters for the iterative graph-partition procedure. Such parameters include the number of partitions or the imbalance factor \cite{schlag2023high}.
\end{itemize}

\subsubsection*{Optimization workflow}
To find the optimal decomposition of the IR, we iteratively partition the circuit until we meet a termination criterion. Fig.~\ref{fig:optimizer} shows an example workflow where we stop when all subcircuits have at most three qubits, the maximum cost is 200, we use the 1q-ops compression method, and our greedy next-leaf function selects the leaf with the most qubits. For graph partitioning, we apply bisection.

We start by compressing the circuit and initializing a contraction tree with a single vertex. In each iteration, we use the next-leaf function to select the leaf for partitioning. For instance, in Fig.~\ref{fig:optimizer}, after step 1, we choose $S_2$ because it has the most qubits. We then partition the selected subgraph into two or more subgraphs, adding new children to the contraction tree. If the partition produces more than two subsets, we build an optimal subtree from the new nodes \cite{gray2021hyper}.
We repeat this process until we exceed the overhead limit or meet the termination criterion and return the IR and contraction tree.

\subsection{Hyperparameter Optimization}
Finding an optimal contraction tree depends heavily on the hyperparameters used in the optimization pipeline (\S~\ref{sec:compiler:optimizer}). We use hyper-optimization to identify the best parameter set, aiming to minimize both PP overhead and estimated error. The optimization is constrained by a maximum PP cost and a user-defined termination state \cite{deb2011multi, akiba2019optuna, gray2021hyper}.

Running the optimizer with different hyperparameters yields varying results in cost and error. The points where no other solution offers lower cost and error are the Pareto-optimal points, forming the Pareto front \cite{emmerich2018tutorial}. These points are optimal solutions for our dual-objective optimization.

To finalize the process, the user supplies a function selecting a point on the Pareto front. By default, we use a function that chooses the point closest to the optimum. This is done by normalizing the points on the Pareto front to $[0, 1]$ for both cost and error, then calculating the Euclidean distance to the ideal point $(0, 0)$, and finally selecting the point with the smallest distance.

% Alternatively, the user can specify a desired trade-off between success and cost, given by the proportion of success divided by cost.
% We then score each point on the Pareto front by the difference between the point's trade-off and the stated trade-off and choose the point with the smallest score.

% \begin{figure}
%     \centering
%     \includegraphics[width=\columnwidth]{figures/benchmarks.pdf}
%     \caption{Circuit ans\"atze used to evaluate \projecttitle{}. \pramod{maybe remove it?}}
%     \label{fig:eval:benches}
% \end{figure}

\subsection{Code Generator}
Once the hyper-optimizer returns an optimal compressed IR and contraction tree, the code generator produces the h-TN.

We generate CTs by examining the cut edges in the \projecttitle{} IR, spanning two subsets determined by the contraction tree's leaves. For each cut edge, we identify its gate type and add the corresponding coefficient tensor $\mathbf{c}$ to the CT set. Wire cuts are handled by adding the two-dimensional tensor from Eq.~\ref{eq:wire-tensor}. Each index is uniquely identified by the operation's index.

A QT represents multiple sub-circuits, each in a different instance. To optimize storage and processing, we generate one blueprint circuit per QT, incorporating single-qubit placeholder gates at the positions of QPD-decomposed gates associated with classical tensors, allowing us to perform circuit optimization only once for each QT. To create a QT from a subset of the IR corresponding to a leaf in the contraction tree, we iterate through the vertices in the subset, adding the relevant gates to the blueprint circuit. For each cut edge, we insert a placeholder gate to insert instance gates during execution. Finally, we add indices to the QT corresponding to the classical tensors for the cut operations acting on the sub-circuit.

\section{The \projecttitle{} Runtime}
\label{sec:runtime}

\begin{figure*}
    \centering    \includegraphics[width=.9\textwidth]{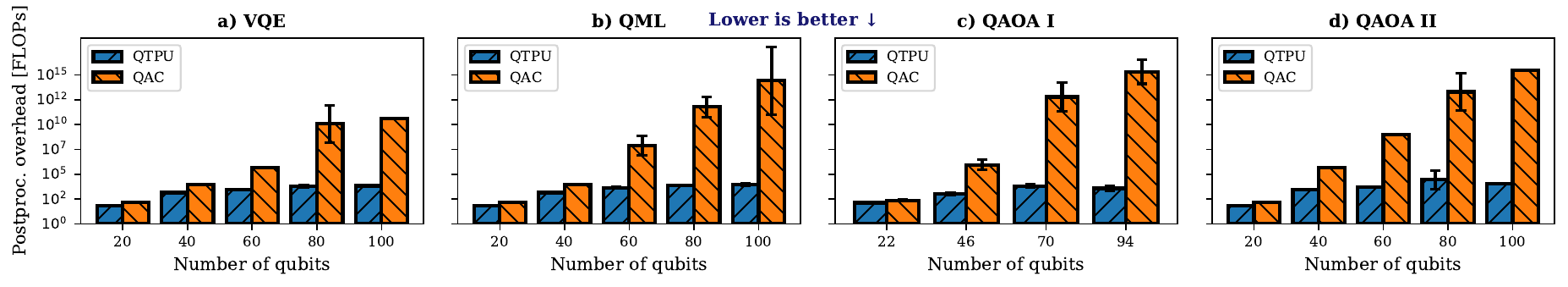}
    \vspace{-3mm}
        \caption{Postprocessing overhead. {\em Classical postprocessing overhead in FLOPs for exact result reconstruction.}}
    \label{fig:eval:cost}
    \vspace{-3mm}
\end{figure*}

We now describe the \projecttitle{} runtime, our scalable system for contracting hybrid TNs.
% As shown in Fig.~\ref{fig:overview}, the runtime consists of two parts: the contraction engine and the QT evaluator.

% \subsection{JIT Low-Rank Approximator}

% Before beginning the actual contraction of a hybrid TN, as a first step in the runtime, the JIT low-rank approximator simplifies the h-TN to a more efficient h-TN at the cost of a small error when applicable.
% We can specify the parameters \textit{tolerance} $\varepsilon$ and the \textit{max to guide the approximation. bond} $b$.
% This ultimately allows us, on the one hand, to bound the error of the approximation to $\varepsilon$\, and, simultaneously, to bound the asymptotical post-processing cost by adjusting $b$.
% \nate{we could put a simple algorithm here that approximates until reaching a given post-processing cost}
% For a fine-grained...
% Finally, we apply the technique described in \S~\ref{sec:idea:approx} to apply chosen the approximation.

\subsection{Contraction Engine}

The contraction engine is the runtime's core component for the contraction of a hybrid TN, running the following steps:

\noindent
\textbf{Step 1 (Simplification and sampling)}: We perform QPD-sampling on the h-TN by first fusing indices through tensor multiplication on CTs between the same QTs and then sampling on the individual CTs a given $s_{\text{QPD}}$ times (\S~\ref{sec:idea:approx}).
This returns a simplified h-TN with truncated CTs and QTs, depending on $s_{\text{QPD}}$. The user can choose the number of samples to enable a tradeoff between accuracy and speed of overall knitting.

\noindent
\textbf{Step 2 (QT evaluation)}: We pass all QTs of the h-TN to the QT evaluator, which returns the respective classical tensor of the expectation values results.

\noindent
\textbf{Step 3 (Contraction preparation)}:
We compute the contraction sequence of the evaluated classical TN using hyperoptimization techniques from previous TN optimization work \cite{gray2021hyper, NVIDIACuTensorNet2024}. Importantly, the contraction sequence of the final classical TN can be computed in parallel with the QT evaluation, as the dimensions of the full TN are already known. This parallelization can significantly reduce the overall computation time for large h-TNs, where both quantum evaluation and contraction-sequence optimizations can be time-consuming.

\noindent
\textbf{Step 4 (Classical contraction)}:
Finally, we allocate and contract the classical TN on the available classical accelerators, using a library for contracting classical tensors \cite{NVIDIACuTensorNet2024, gray2021hyper}.

\subsection{Quantum Tensor Evaluator}
The QT evaluator is responsible for transforming QTs into CTs corresponding to the results of each subcircuit instance.
A QT consists of a blueprint circuit with placeholder gates and a tensor of gates to be inserted into the blueprint circuit.

For each QT to evaluate, we first apply the circuit optimization to the blueprint circuit to perform the extensive optimization only once, meaning that each circuit instance is optimized from one circuit optimization.
Then, we generate the circuit instances by inserting the corresponding gates of each element in the instance tensor into the optimized blueprint circuit.
We then pass the resulting quantum circuits and the number of shots required by the specific QPD decomposition to an interface capable of executing the quantum circuits and returning the samples.
Such an interface could be a single QPU or a cluster of QPUs and/or simulators that can execute the circuits in parallel \cite{seitz2024scim, giortamis2024qos}.
Finally, we construct and return a classical tensor from the results with the same dimensions as the respective QT.

\begin{figure*}
    \centering    \includegraphics[width=.9\textwidth]{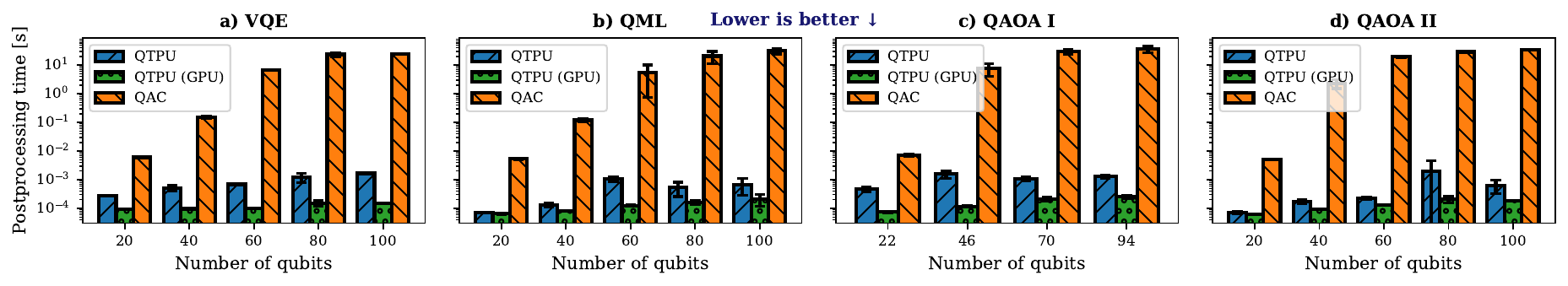}
     \vspace{-3mm}
        \caption{Postprocessing time. {\em Classical postprocessing time for exact result reconstruction of \projecttitle{} on CPU and GPU vs. approximate reconstruction using QAC. We limit QAC at $10^5$ QPD-samples, as it would otherwise not terminate in a reasonable time.}}
    \label{fig:eval:runtime-1}
    \vspace{-3mm}
\end{figure*}

\section{Evaluation}

% \begin{figure*}
%     \centering
%     \includegraphics[width=\textwidth]{figures/comp_improve.pdf}
%     \caption{Impact of the \projecttitle{}'s Compiler. {\em (a) relative circuit depth, (b) relative number of non-local gates, and (c) relative estimated success probability}}
%     \label{fig:eval:compiler}
% \end{figure*}

\subsection{Methodology}

\myparagraph{\projecttitle{} implementation}
We implement \projecttitle{} in Python v3.11.9 building on Qiskit v1.1.0  for representing quantum circuits and the Qiskit Addon Cutting (QAC) v0.7.3 for the QPD decompositions of gates.
For the compiler, we build on cotengra \cite{gray2021hyper} and quimb \cite{gray2018quimb} for the representations of the IR, contraction trees and hybrid TNs, kahypar \cite{DBLP:phd/dnb/Schlag20} for efficient graph partitioning and optuna for hyperparameter optimization \cite{akiba2019optuna}.
We implement the \projecttitle{} runtime using quimb for contractions on CPUs and cuTensorNet on GPUs \cite{gray2018quimb, NVIDIACuTensorNet2024}.

\myparagraph{Framework configuration}
Unless otherwise noted, we use the \projecttitle{} to compile the circuits into h-TNs with subcircuits with maximal 15 qubits.
When running circuit simulations, we use a \textit{BackendSamplerV2} primitive on top of a GPU-accelerated \textit{AerSimulator} performing 20,000 shots per circuit.

\myparagraph{Experimental setup}
We conduct our experiments on an Intel Xeon Platinum 8360Y CPU with 72 cores (144 HT), 2.40GHz, 362GB RAM, and an NVIDIA A100 80GB PCIe GPU.

% We conduct three types of experiments: (1) circuit transpilation with and without \projecttitle{}'s compiler to measure the circuit's properties post-compilation, (2) runs on real QPUs for measuring the circuit's fidelity, and (3) classical simulation of large circuits cut into fragments of different sizes.
% For (2) we conduct our experiments on Falcon r5.11H QPUs, namely the 7-qubit IBM Perth and the 27-qubit IBMQ Kolkata.
% %(Fig. \ref{fig:qpus}). 
% For (1) and (3) we use the Qiskit Transpiler and Qiskit Aer \cite{qiskit-transpiler, qiskit-aer}, respectively, and run on our local classical machines.  For classical tasks, i.e., transpilation, post-processing (knitting), and simulation, we use a server with a 64-core AMD EPYC 7713P processor and 512 GB ECC memory.

% \subsubsection{Configuration} We use the \textit{Qiskit} \cite{Qiskit} Python SDK version 0.41.0 for quantum circuits and simulations. We transpile any quantum circuit we run with the highest optimization level.

% To get a meaningful measurement of the fidelity or circuit properties on real QPUs, we run QVM only on a single QPU.
% We utilize every system core when we benchmark the performance of the \projecttitle{} runtime with simulators.

\myparagraph{Benchmarks}
We use four commonly used ansätze as benchmarks: (a) Variational Quantum Eigensolver (VQE) with an efficient $SU(2)$ ansatz and linear entanglement, (b) Quantum Machine Learning (QML) with a $ZZ$-feature map and pairwise entanglement, (c) Quantum Approximate Optimization Algorithm (QAOA-I) for a random clustered graph with 70\% intra-cluster and 30\% inter-cluster connection probabilities \cite{lowe2023fast}, and (d) QAOA-II for a random clustered graph with 70\% intra-cluster connections and $p$ edges between adjacent clusters \cite{perlin2021quantum}.
% We chose these for benchmark purposely, as VQE and QAOA-I favor wire cuts, while QML and QAOA-II favor gate cuts due to their gate structures.

\myparagraph{Baselines}
We use the current state-of-the-art circuit-knitting framework, the ``Qiskit Addon Cutting'' (QAC) v0.7.3, as our main baseline \cite{circuit-knitting-toolbox}. QAC is the only existing QPD-based knitting framework that, like \projecttitle{}, allows for both wire and gate cutting.
We also compare against classical GPU-accelerated TN simulation cuTensorNet \cite{NVIDIACuTensorNet2024}.

\myparagraph{Metrics} We evaluate \projecttitle{} based on the following metrics.

\begin{enumerate}
    \item \textit{Postprocessing (PP) overhead}: The number of float operations (FLOPs) required for classical PP.
    % \item \textit{Number of QPD-samples}: Number of sam% The QPD-sampling overhead for a circuit. Used to compare our compiler to CKT's circuit cutter to determine which finds a better cut-solution.
    % \item \textit{Runtime}: Runtime of compiling, running, or post-processing.
    \item \textit{Runtime}: Runtime in seconds for either (1) classical PP or (2) full knitting, including QT-evaluation. 
    % We mainly compare these times to (1) PP or (2) full knitting in QAC.
    \item \textit{Number of two-qubit gates}: The relative number of two-local gates of the circuit (lower is better). If a circuit contains multiple subcircuits, we report the maximum of all subcircuits.
    \item \textit{Estimated Error}: Estimated error of running a circuit, assuming a uniform error model of $10^{-4}$ and $10^{-3}$ for single- and two-qubit gates, respectively \cite{IBMQuantum2024}. For multiple subcircuits, we provide the maximal error.
    % \item \textit{Error}: The absolute error of estimating an expectation value. We estimate the expectation value of the observable $O = Z^{\otimes n}$ for simplicity.
    % The fidelity of running a circuit on a noisy (QPU). For reconstruction of the full probability distribution, we use the Hellinge fidelity \cite{hellinger1909neue, fidelity-qiskit}; for expectation values, we use the absolute error.
\end{enumerate}

\subsection{Postprocessing Analysis}
\label{sec:eval:cost}

To show \projecttitle{}'s capability to mitigate exponential PP overhead, we use our compiler to compile benchmarks ranging from 20 to 100 qubits, reducing them to subcircuits of no more than 15 qubits, requiring 2 to 10 cuts.
We then measure both the PP overhead in FLOPs and the total PP time.
We pose the following research questions (RQs).

\myparagraph{--- RQ1: Postprocessing overhead}
\textit{How does \projecttitle{}'s approach impact the PP overhead of circuit knitting?}

\noindent
First, we compute the PP overhead, assuming all QPD coefficients are sampled, and compare this with the PP overhead used by QAC (see Fig.~\ref{fig:eval:cost}).
We observe only a linear overhead increase for \projecttitle{}'s PP overhead as \projecttitle{} allows accelerated tensor-based PP. At the same time, we see a high exponential growth in overhead for the PP of QAC.
In total, qTPU improves overhead by $10^{15} \times$ on average and up to $10^{17}\times$.

\myparagraph{RQ1 takeaway}
\projecttitle{} reduces the PP overhead by orders of magnitude compared to state-of-the-art knitting as it avoids exponential overheads through its h-TN contraction.

\myparagraph{--- RQ2: Postprocessing time}
\textit{By how much can \projecttitle{} reduce PP time?}

\noindent
Correlated with the reduction in overhead, we see a significant speedup in PP time compared to QAC.
We need to run QAC with an approximation using $10^5$ QPD-samples, as it would otherwise not terminate for more than $\sim$50 qubits.
Despite only approximating the result with QAC, our results, shown in Fig.~\ref{fig:eval:runtime-1}, indicate an average speedup of $10^{4.3}\times$. 
% when conducting PP on a CPU.

As we can offload the PP onto a GPU, we can obtain a further speedup by up to $18.5 \times$ compared to qTPU's CPU PP.
Importantly, even when the theoretical overhead is the same, e.g., for the 20-qubit benchmarks, \projecttitle{} still achieves a speedup of $\sim 57\times$ through its tensor processing.

\myparagraph{RQ2 takeaway}
\projecttitle{}'s enables a speedup in PP of $10^{4.3}\times$ on average compared to QAC.
Since \projecttitle{}'s PP consists of tensor contractions, we can offload the process to the GPU, allowing a further speedup of up to $18.5 \times$.

\begin{figure*}
    \centering    \includegraphics[width=.9\textwidth]{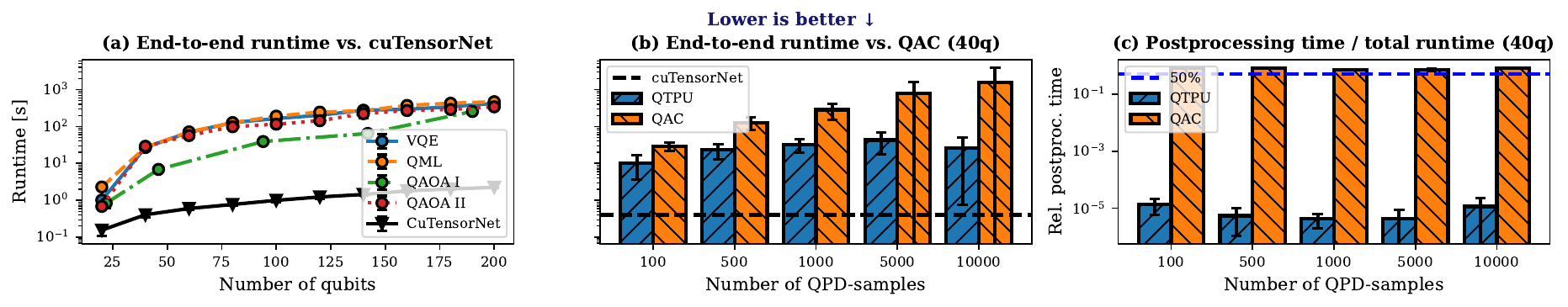}
        \vspace{-3mm}
        \caption{End-to-end runtime. {\em (a) End-to-end h-TN contraction time using a 15-qubit statevector simulator QPU vs. TN-simulation. (b) End-to-end h-TN contraction time depending on the number of QPD-samples vs. QAC knitting. (c) Portion of time spent on postprocessing relative to full runtime.}}
    \label{fig:eval:end2end}
    \vspace{-3mm}
\end{figure*}

\begin{figure}
    \centering    \includegraphics[width=\columnwidth]{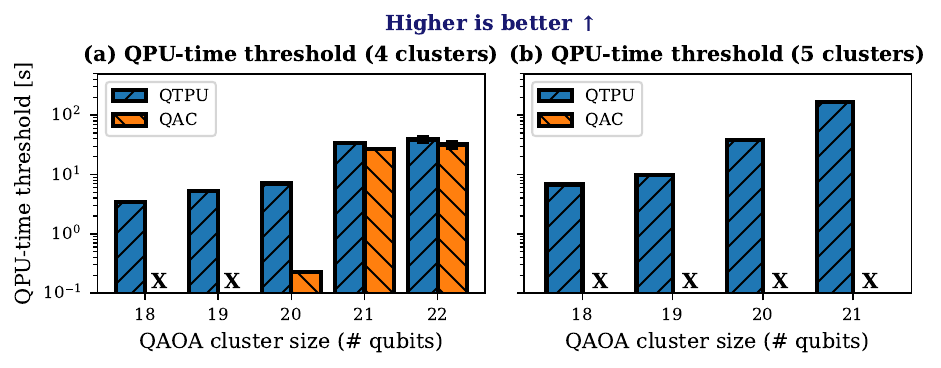}
    \vspace{-3mm}
        \caption{Runtime threshold analysis. {\em The maximal allowed time to spend on circuit evaluation during knitting to outperform classical TN-simulation. An $X$ indicates that postprocessing exceeds full TN-simulation time.}}
    \label{fig:eval:threshold}
\end{figure}

\subsection{End-to-end Runtime Analysis}

To show the practicality of \projecttitle{} as a valuable method for large-scale hybrid computations, we need to analyze its end-to-end runtime and its ability to scale beyond purely classical simulation techniques by using QPUs and GPUs.
To this end, we examine the runtime of \projecttitle{} to contract an h-TN using a 15-qubit state vector simulator as a mock QPU and compare it to QAC and cuTensorNet.

\myparagraph{--- RQ3: End-to-end scalability}
\textit{How does \projecttitle{}'s end-to-end runtime scale for large circuits?}

\noindent
First, to demonstrate \projecttitle{}'s scalability, we conduct end-to-end simulations of benchmarks from 20 to 200 qubits, as shown in Fig.~\ref{fig:eval:end2end} (a). \projecttitle{} scales linearly with the number of qubits and subcircuits, taking $\sim$30 seconds for the 40-qubit benchmark with three subcircuits and $\sim$400 seconds for the 200-qubit benchmark with more than 12 subcircuits.

The TN-simulator based on cuTensorNet achieves sub-second runtimes by running benchmarks directly in GPU memory. In contrast, \projecttitle{} simulates hundreds of subcircuits sequentially during h-TN contraction, though this could be improved by using real QPU(s) to run subcircuits faster.

Despite this, \projecttitle{} shows comparable scalability to efficient TN-contraction-based simulation techniques, enabling the execution of circuits that cannot be run with QAC or state vector simulation alone due to their exponential complexity.

\myparagraph{RQ3 takeaway}
\projecttitle{} achieves linear scaling for the end-to-end simulation of large quantum circuits that would not be possible to run with a single state-vector simulator or QAC.

\myparagraph{--- RQ4: Runtime-dependence on QPD-sampling}
\textit{How does end-to-end runtime behave depending on the number of QPD-samples compared to QAC?}

\noindent
To reduce runtime during knitting, we can perform QPD-sampling on the coefficients to approximate the result, in return for losing accuracy as we use fewer samples (\S~\ref{sec:idea:approx}).
To show how sampling affects overall runtime scaling compared to QAC, we run our benchmarks of 40 qubits and show the average end-to-end runtime over all benchmarks in Fig.~\ref{fig:eval:end2end} (b).

We observe an exponential increase in runtime for QAC as we increase the number of QPD samples, as it directly depends on the number of global coefficients sampled.
However, in \projecttitle{}, since we sample directly in the efficient h-TN form, QPD-sampling only slightly affects runtime as the number of samples increases exponentially.
By leveraging its efficient form, \projecttitle{} can still speed up overall runtime by orders of magnitude with an average of $20.7\times$ and up to $36.3\times$.

\myparagraph{RQ4 takeaway}
As \projecttitle{} leverages its efficient h-TN form during QPD sampling, its runtime is comparatively unaffected by the total number of QPD samples. It enables a significant speedup over QAC by $20.7\times$ on average.

\myparagraph{--- RQ5: Bottleneck analysis}
\textit{Where does the main bottleneck in end-to-end runtime lie in \projecttitle{} compared to QAC?}

\noindent
To investigate entirely where our approach's bottleneck compared to QAC is, we analyze the relative time spent on PP compared to the end-to-end runtime.
We show the results in Fig.~\ref{fig:eval:end2end} (c), using the same benchmarks as the previous result.
In QAC, we observe that classical PP dominates the overall runtime of knitting, with 75.7\% on average and up to 79.5\%. 
For \projecttitle{}, however, PP only occupies a fraction of $10^{-3}$\% on average of overall runtime, virtually eliminating the imposed bottleneck of PP. Therefore, the bottleneck of circuit knitting is only the evaluation of subcircuits on QPUs, where the sampling overhead remains unchanged.

\myparagraph{RQ5 takeaway}
\projecttitle{} allows us to eliminate the PP-bottleneck of circuit knitting, as it now only occupies a fraction of the total circuit knitting time.

\begin{figure*}
    \centering    \includegraphics[width=.95\textwidth]{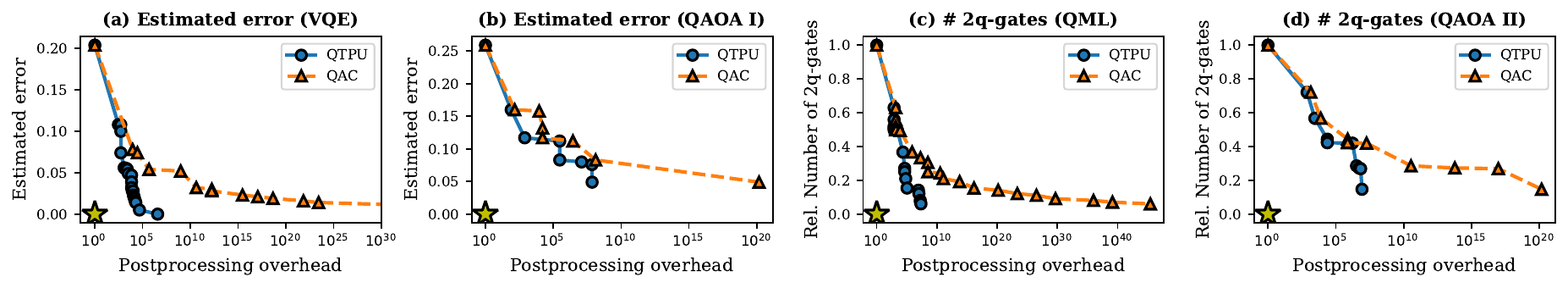}
        \vspace{-3mm}
        \caption{Compilation tradeoff. {\em Pareto-optimal circuit compilations of 100-qubit benchmarks explored by \projecttitle{}'s compiler.}}
        \vspace{-3mm}
    \label{fig:eval:tradeoff}
\end{figure*}

\begin{figure}
    \centering    \includegraphics[width=\columnwidth]{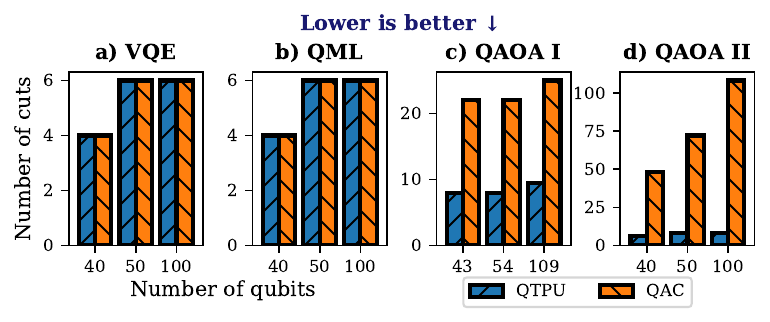}
        \caption{Compiler Optimality. {\em Number of cuts after compiling a circuit to have $\sim$25\% of the original number of qubits.}}
    \label{fig:eval:compiler_cuts}
\end{figure}

\myparagraph{--- RQ6: Outperforming classical simulation}
\textit{In which scenarios can \projecttitle{}'s hybrid execution enable outperforming classical TN-simulation?}

\noindent
We demonstrate that \projecttitle{} enables the execution of large circuits that cannot be efficiently run with TN-based techniques. Using QAOA II benchmarks with increasing cluster sizes makes them increasingly difficult to simulate due to memory constraints. We conduct benchmarks on four and five QAOA clusters, ranging from 18 to 22 qubits, measuring the simulation time with cuTensorNet and the postprocessing (PP) time for \projecttitle{} and QAC. The differences in TN-simulation time and PP are shown in Fig.~\ref{fig:eval:threshold}, indicating the maximum time available for subcircuit evaluation to outperform cuTensorNet.

We find that \projecttitle{} sets a threshold runtime for QPUs between 3.38 seconds and 168.7 seconds, which increases rapidly as cluster sizes complicate TN simulation. This provides a realistic opportunity to surpass TN simulation as the number of clusters increases since QPU runtime depends solely on circuit depth. However, reaching this threshold is significantly more difficult for QAC's PP, as it takes, on average, 9.2 times longer than the full TN simulation, hindering any chance of surpassing simulation through knitting.

\myparagraph{RQ6 takeaway}
For clustered benchmarks that become difficult to simulate classically, \projecttitle{}'s hybrid processing allows a realistic chance to outperform classical simulation because of its negligible PP time.

\subsection{Compiler Analysis}
\label{sec:eval:compiler}

Finally, we discuss the performance of our compiler.
In doing so, we focus on its novelty in enabling us to explore a tradeoff between expected error and overhead. Finally, we also compare the compiler against QAC's automatic cut-finder.
% to determine optimal cuts of quantum circuits.

\myparagraph{--- RQ7: Error-overhead tradeoff}
\textit{How does the \projecttitle{} compiler enable an optimal trade-off between error and overhead?}

The \projecttitle{} compiler employs hyper-parameter optimization on various \projecttitle{}-specific parameters to compute circuit decompositions, each representing a tradeoff between estimated error in subcircuits and PP cost. To showcase this, we run our hyper-optimizer with 200 trials for 100-qubit benchmarks and record the Pareto-optimal solutions. For example, we use (1) the estimated error and (2) the relative number of two-qubit gates as error measures. The results are shown in Fig.~\ref{fig:eval:tradeoff}, including the respective overheads of QAC.

For the simpler benchmarks of VQE and QML, we observe a steep Pareto front. Notably, the \projecttitle{} compiler provides solutions with nearly zero estimated error by cutting the circuit into subcircuits of only a few gates, enabled by the linearly increasing overhead, reaching a maximum of approximately $\mathcal{O}(10^8)$ FLOPs, allowing most of the circuit to run classically.

In contrast, for the more complex clustered QAOA benchmarks, we observe a flatter curve, as more cuts are needed to achieve low estimated error, making it difficult to reach minimal errors with reasonable overhead.

The corresponding curves of QAC's cost show an exponential increase in PP cost as we attempt to reduce expected error, closely following the previous results in \S~\ref{sec:eval:cost}. This enables \projecttitle{} to come significantly closer to the optimum (yellow star).

\myparagraph{RQ7 takeaway}
\projecttitle{}'s compiler facilitates the exploration of the trade-off space between PP cost and user-defined error estimation, significantly improving on previous work focusing solely on minimizing overhead. During its exploration, the optimizer even enables exponential error reduction with only a linear increase in overhead.
% enabling it to come $\times$ closer to the theoretical optimum than QAC.
% Thus, \projecttitle{}'s compiler paves the way for highly customizable quantum-classical PP.

\myparagraph{--- RQ8: Compiler efficiency}
\textit{How does \projecttitle{}'s compiler compare to QAC's automatic cut-finder to determine efficient cuts?}

\noindent
To answer this question, we compile benchmarks with 40-100 qubits, cutting the circuit into subcircuits that are $\sim$ 25\% of the original size. We present the results, showing the number of cuts in the compiler's solution in Fig.~\ref{fig:eval:compiler_cuts}, and compare them to QAC's automatic cutter \cite{shehzad_automated_2024}. 
We observe that the simpler benchmarks, such as VQE and QML, perform comparably to QAC. However, as the benchmarks increase in complexity, particularly with highly clustered interconnectivity of gates like in QAOA, our hyperparameter optimization enables us to identify efficient cuts where QAC struggles.

\myparagraph{RQ8 Takeaway}
\projecttitle{}'s compiler can efficiently cut complex quantum circuits where single graph-partitioning methods fail. At worst, \projecttitle{} is on par with existing methods.

\section{Related Work}

% qTPU enables to combine TN methods and quantum computing, allowing the use of either them with depending on which device offers an 

% \pramod{slightly shorten it?}

\myparagraph{Quantum circuit knitting}
The Qiskit Addon Cutting (QAC) serves as the current de facto framework for circuit knitting, integrating prior work on quantum gate decomposition (QPD) and circuit cutting \cite{tang2022scaleqc, mitarai2021constructing, peng2020simulating}, and, like \projecttitle{}, it supports both gate and wire cuts. Ongoing research aims to reduce sampling overhead by identifying lower-overhead decompositions for multi-qubit gates or enabling classical communication between quantum processing units (QPUs) \cite{brenner2023optimal, schmitt2023cutting, ufrecht2023cutting, ufrecht2023optimaljoint, lowe2023fast}; in contrast, we focus on minimizing postprocessing (PP) overhead while enabling large-scale hybrid processing with classical accelerators. Other studies explore efficient circuit cuts, targeting either wire or gate cuts \cite{tang2021cutqc, cambiucci2023hypergraphic} or both simultaneously \cite{shehzad_automated_2024, brandhofer2023optimal}, relying on graph partitioning routines designed to fit circuits to specific QPU sizes. Our work facilitates scalable wire and gate cutting while optimizing the trade-off between error and overhead through hyperparameter tuning and efficient TN optimization. Recent studies have explored using TNs to speed up PP \cite{tang2022scaleqc, ren2024hardware}, focusing only on wire or gate cutting. In contrast, we present a holistic approach through a compiler that optimizes for h-TN processing.

\myparagraph{Tensor network methods}
\projecttitle{} draws inspiration from well-studied TN methods, establishing connections between techniques used in classical TN optimization and those implemented in our framework. In particular, the optimizer within the \projecttitle{} compiler resembles the standard approach for determining an optimized contraction sequence of a classical TN, making it highly suitable for optimizing our hybrid circuit contraction \cite{gray2021hyper, NVIDIACuTensorNet2024, schindler2020algorithms}. The compression of our IR parallels rank-simplification optimization while sampling on a hybrid TN can be likened to low-rank approximations \cite{bridgeman2017hand}. Furthermore, prior research has introduced the concept of a "hybrid TN" for performing quantum ground state optimizations \cite{yuan2021quantum, schuhmacher2024hybrid}, which focuses on offloading parts of an iterative TN algorithm onto a quantum processing unit (QPU), presenting an approach that contrasts with ours.

\myparagraph{Hybrid computing}
A related branch of research focuses on architectures for scheduling and running complex quantum-classical algorithms in hybrid clusters \cite{chen2024quantum, giortamis2024orchestrating, giortamis2024qos, salm2022prioritization, wang2024qoncord, schulztowards}. Our hybrid circuit contraction enforces a straightforward processing model, requiring only a simple execution model to execute subcircuits on QPUs. However, we can easily expand our runtime by integrating it with sophisticated schedulers \cite{kaewpuang2023stochastic, seitz2024scim, ravi2021adaptive}.
\section{Conclusion}
Circuit knitting holds promise for scaling quantum computing, but its exponential brute-force postprocessing limits its potential. To address this, we introduced \projecttitle{}, a scalable framework for hybrid circuit contraction, representing quantum circuits as hybrid tensor networks (h-TNs) to accelerate circuit knitting. Our \projecttitle{} compiler generates hyper-optimized h-TNs, balancing error and overhead, while the runtime efficiently contracts these networks on GPUs and QPUs. Evaluation shows \projecttitle{} significantly improves postprocessing and runtime by $10^4\times$ and $20.7 \times$ on average, respectively. We believe our approach is a crucial step toward making hybrid quantum-classical processing useful.
% viable and superior to purely classical or quantum methods.

\myparagraph{Artifact} \projecttitle{} will be publicly available.

% In future work, we must figure out which scenarios circuit cutting is suited for, and in which cases only tensor network contraction might be the better choice

\section*{Acknowledgements}
We thank Lukas Schmitt and Minh Chung for insightful discussions. We thank the Quantum Computing and Technologies (QCT) and Future-Compute group at Leibniz Supercomputing Centre for their support. Funded by the Bavarian State Ministry of Science
and the Arts as part of the Munich Quantum Valley (MQV).

\bibliographystyle{unsrt}
\bibliography{references}

\end{document}